# IMPACT OF EFFECTUAL PROPENSITY ON ENTREPRENEURIAL INTENTION

Alicia Martín-Navarro [a,b], Felix Velicia-Martín [a], José Aurelio Medina-Garrido [b,*], Pedro R. Palos- Sánchez [a]

[a] *University of Seville, Spain*
[b] *INDESS, University of Cadiz, Spain*
*\* Corresponding author*

This version was submitted for evaluation in the "Journal of Business Research" and accepted for publication after the review process. The final published version can be found at:
https://doi.org/10.1016/j.jbusres.2022.113604
We acknowledge that Inderscience holds the copyright of the final version of this work.
Please, cite this paper in this way:
Martín-Navarro, A., Velicia-Martín, F., Medina-Garrido, J. A., & Palos-Sánchez, P. R. (2023). Impact of effectual propensity on entrepreneurial intention. *Journal of Business Research*, 157(January).

**Abstract**
For decades, entrepreneurship has been promoted in academia and tourism sector and has it been seen as an opportunity for new business ventures. In terms of entrepreneurial behavior, effectual logic shows how the individual uses his or her resources to create new opportunities. In this context, this paper aims to determine effectual propensity as an antecedent of entrepreneurial intentions. For this purpose, and based on the TPB model, we conducted our research with tourism students from Cadiz and Seville (Spain) universities with Smart PLS 3. The results show that effectual propensity influences entrepreneurial intentions and that attitude and perceived behavioral control mediate between subjective norms and intentions. Our research has a great added value since we have studied for the first time efficacious propensity as an antecedent of intentions in people who have never been entrepreneurs.

## 1. Introduction

The academic literature highlights the importance of entrepreneurial intention as a precursor to entrepreneurial decision making (Maheshwari, 2021). Fostering entrepreneurship is crucial in all countries, as it contributes to the development of the economy by increasing productivity, creating new employment opportunities, and revitalizing markets through the development of new products (Esfandiar et al., 2019). In particular, the tourism sector offers many entrepreneurship and business development opportunities. This industry promotes economic development in its geographical area of influence, as it acts as an engine for other economic sectors. In this context, the tourism entrepreneur is key to developing attractive destinations, creating employment, and increasing labor productivity and income of the local population. (Tleuberdinova et al., 2021)

In the initial stages of the entrepreneurial process, the decision to entrepreneurship is preceded by the detection of opportunities and the intention of creating a new venture (Liñán, 2008). The Theory of Planned Behavior (TPB) (Ajzen et al., 1991) proposes a helpful model for analyzing entrepreneurial intention. (Krueger et al., 2000; Liñán, 2004; Soomro et al., 2018; Urban & Ratsimanetrimanana, 2019). Applying TPB concepts, entrepreneurial intention depends on the personal attitude (PA), subjective norms (SN) and perceived behavioral control (PBC) of the potential entrepreneur. In another sense, and according to Sarasvathy (2001), the entrepreneur may apply a causal or effectual logic while creating his or her new business. To date, scholars have analyzed the effectual versus causal logic by studying the behavior of individuals who have already created companies (Brettel et al., 2012; Guo et al., 2016; Schmidt & Heidenreich, 2014). However, efficacy research does not analyze entrepreneurial intention. (Shahidi, 2020). The literature has not studied the individual propensity toward one or the other logic in the case of the potential entrepreneur, the one who intends to become an entrepreneur but has not yet done so. The theories of entrepreneurship concerning the study of entrepreneurial intention seemed incompatible with the theories on the type of entrepreneurial behavior to develop since they apply to two different moments of the entrepreneurial process. Considering this gap, the study of the propensity for effectual or causal behavior prior to the firm's creation would add value to entrepreneurship research. (Martín-Navarro et al., 2021). Based on the above, the aim of this paper is to test the direct and indirect impact of the effectual propensity of potential entrepreneurs on their entrepreneurial intention.

The literature on entrepreneurship widely recognizes the university's role as a pool of entrepreneurs. (Lechuga Sancho et al., 2021). The reasons why students do or do not intend to run their own business have been the subject of interest in the entrepreneurship literature (Trang & Doanh, 2019). Because the tourism sector offers many opportunities for entrepreneurial development, our study was conducted among the University of Cadiz and the University of Seville tourism students. Exploring the entrepreneurial intentions of this type of student is crucial for the current and future development of the tourism and hospitality industry. (Tsai et al., 2016).

The impact of effectual propensity on entrepreneurial intention modeled according to the TPB was analyzed to achieve the objective. Hypothesis testing revealed that seven hypotheses were statistically significant, and two were rejected. Expressly, it was rejected that SNs directly impact entrepreneurial intentions and perceived behavioral control. However, the effectual propensity of the potential entrepreneur did show a direct, even slightly, effect on intentions and indirect effects through PA and PBC.

## 2. Background and Research Hypothesis

The literature on entrepreneurship provides adequate justification for the relationships to test in this study. The following is a brief description of the basic concepts related to effectual behavior and TPB application to entrepreneurial intention models. Those concepts will subsequently allow us to justify and establish the hypotheses to test to respond to our research objective.

### 2.1. Effectuation

Decisions on how to develop a business idea, acquire the necessary resources and implement effective decision making occur under conditions of uncertainty (Villani et al., 2018). For a long time, entrepreneurship researchers have assumed that individuals pursuing entrepreneurial opportunities are guided solely by rational, goal-directed behavior (Perry et al., 2012). This thinking fits into a causal behavioral model of entrepreneurship (Sarasvathy, 2001). In contrast, the effectual behavior theory of entrepreneurship proposes a different approach to explain how some entrepreneurs make decisions (Sarasvathy, 2001). Thus, causation and effectuation represent two distinct frameworks applicable to new venture creation. The first is characterized by careful planning, while the latter is based on a more flexible, adaptive and experimental strategy. (Read & Sarasvathy, 2005).

Focusing on the above framework, the logic of effectuation is that entrepreneurs use a set of means already given to them and focus on deciding what they can create with them (Sarasvathy, 2001). When creating new ventures, entrepreneurs who follow an effectual approach, as they make decisions and observe the results of those decisions, use this new information to change course.

Thus, entrepreneurs who follow an effectual logic are less likely to plan for the future and prefer to change their initial goals and vision for their new venture. Thus, instead of predicting the future, they are more likely to work with the means they have under their control and make the necessary adjustments. (Dew et al., 2009).

For Sarasvathy (2001), this approach offers five decision-making principles (means, association, affordable loss, contingencies, and control): (1) start with the *means* (entrepreneur's skills, knowledge and social relationships), wherein the process of creating a new venture, goals emerge based on the entrepreneur's available *means*; (2) form partnerships to seek business opportunities together, share resources and work together, with less uncertainty and more control of each partner's business; (3) focusing on downside risk, and worrying more about the *affordable loss* than the profits to achieve; (4) taking advantage of contingencies and not thinking of these contingencies as obstacles, but as new opportunities to exploit; and, (5) controlling the future, which is uncertain but controllable because entrepreneurs can influence trends, create new markets and face new challenges.

To understand the behavior of this type of individual, we must determine their orientation towards effectual principles. Sarasvathy's (2001) theory has been widely studied by the academic community as applied to entrepreneurs starting new ventures (Burmeister-lamp, 2016; Chandler et al., 2011; Matalamäki et al., 2017). However, few have been concerned with the effectual orientation of potential entrepreneurs. In this regard, Martín-Navarro et al. (2021) are the first to use the concept of effectual propensity (EP). The EP determines the orientation of a person towards an effectual logic before starting a business.

## 2.2. Theory of Planned Behavior (TPB)

Entrepreneurial intention has become a fascinating field in entrepreneurship research. It becomes a way to find out the desire and commitment of people to create new ventures. Entrepreneurship research shows those personal and contextual factors that motivate the individual to start an entrepreneurial project (Shahzad et al., 2021). To this end, the Theory of Planned Behavior (TPB), an accepted model for analyzing the propensity of individuals to develop particular behavior, helps study entrepreneurial intention. (Krueger et al., 2000). According to the TPB model, intentions are shown to be the result of three factors: (1) personal attitude (PA) toward a given behavior, (2) subjective norms (SN) determined by social pressure and, (3) perceived behavioral control (PBC) understood as the ability to carry out a specific behavior. (Ajzen, 1991). This theory has been widely used in entrepreneurship and has been empirically tested by numerous scholars. (Liñan, 2004; Soomro et al., 2018; Urban & Ratsimanetrimanana, 2019). In this sense, the literature on entrepreneurship adapts the general elements of the TPB to the explanation of entrepreneurial intention. Thus, PA toward behavior refers to how individuals have a positive or negative personal appraisal of being an entrepreneur. Individuals' attraction toward the desire to start their own business affects their entrepreneurial intention. (Ajzen, 1991). SN reflects the pressure individuals receive from their social environment to perform certain behaviors. (Ajzen, 1991). In an entrepreneurial setting, SN refers to the importance that individuals attach to the approval of their closest circle (family, friends or other people of reference) to their decision to start a business. (Liñán, 2008). Finally, PBC refers to the perception that individuals have that entrepreneurship is easy or difficult and can control this behavior. (Ajzen, 2002).

## 2.3. Research model

Attitudes are psychological traits that influence entrepreneurial intentions. Attitudes can change over time and be influenced by education or experience. Underlying attitudes are cognitive structures. (Krueger, 2007). The cognitive structure is the set of ideas that an individual has about a given area of knowledge and how he or she organizes them in his or her mind. We can point to effectual logic among the cognitive structures, including decision-making and problem-solving methods. (Sarasvathy & Dew, 2008). Thus, an individual's EP can be a precedent for PA and can be raised:

**H1:** PE is positively related to PA toward entrepreneurship.

According to Yoon & Cho (2021), an internal locus of control is positively related to effectual and causal behavior. However, the central theme of the entrepreneur's effectual logic is control and not prediction, which is characteristic of causal behavior. (Sarasvathy, 2001). Effectual behavior considers that individuals perceive that they can control their future. Even though the future is uncertain, effectual entrepreneurs perceive creating new markets and opportunities. (Sarasvathy, 2001). This idea is a perception of having the ability to control events. Therefore, control as an effectual principle is connected to the TPB concept of PBC, understood as the perception that one can carry out a particular behavior. (Ajzen, 1991). For this reason, it can be argued that:

**H2:** PE is positively related to PBC.

As discussed, according to efficacy, effectual behavior relies on five principles: means, partnership, affordable loss, contingencies, and control orientation. Traditionally, the entrepreneurship literature analyzes effectual behavior following the creation or development of a firm. Following Martín-Navarro et al. (2021), we will apply these same principles to the effectual propensity of individuals, i.e., the tendency to develop effectual behavior before creating the firm. A review of previous literature allows us to relate the principles of effectual behavior and the analogy of the propensity towards effectual behavior to entrepreneurial intention. Thus, as part of his or her means, knowledge and skills of the potential entrepreneur have a positive effect on entrepreneurial intention (Liñan, 2004). Similarly, risk propensity, which we can relate to *affordable loss*, also influences entrepreneurial intention. (Shahzad et al., 2021). Concerning the effectual principle of *partnership*, there is evidence of how collective entrepreneurship explains entrepreneurial intention among rural youth members of agricultural cooperatives. (Bouichou et al., 2021). Along the same lines, there is also evidence that team cooperation moderates the positive relationship between entrepreneurial education and entrepreneurial intention. (Li & Wu, 2019). As the ability to recognize opportunities in difficult situations, contingency can also be related to intentions. Thus, Al Mamun et al. (2016) found that students who could identify opportunities increased entrepreneurial intentions. Finally, environmental *control* orientation is closely related to the internal locus of control concept. (Somi, 2015). This concept assumes that the individual has control over his or her destiny and thinks that what happens depends on his or her abilities and knowledge. (Mueller & Thomas, 2001). Therefore, if the entrepreneur feels strong in controlling the destiny, he or she will be more confident about the chances of success, thus increasing his or her entrepreneurial intentions. In this sense, many scholars have found the effect of internal locus of control on entrepreneurial intentions. (Annisa et al., 2021; Erickson & Laing, 2016; Nasip et al., 2017; Torres et al., 2017). Given that EP is governed by the same principles as effectual behavior after entrepreneurship (Martín-Navarro et al., 2021), we can affirm that there must also be a relationship between PE and EI. There is empirical evidence for this relationship. In a sample of graduates from a leading business school in Pakistan, it was found that training students in the effectual entrepreneurial logic significantly increased their intentions to start a business. (Qureshi & Mahdi, 2014). Based on the above, we can argue that:

**H3:** PE is positively related to entrepreneurial intention.

On the other hand, SN involves a significant social influence on individuals. When individuals are in a group, they feel that they must have appropriate behavior, usually conditioned by family, friends or peers (Yasa et al., 2021). In this way, SN presents a significant relationship with the individual's attitude (Kotler & Keller, 2015). Piroth et al. (2020) found empirical evidence for the positive effect of SN on PA in a study on online grocery shoppers. Similarly, Yasa et al. (2021) tested how SN had a positive impact on PA wearing a face mask. These arguments allow the following hypothesis:

**H4:** SN is positively related to PA toward entrepreneurship.

The support of the most influential people who make up the SN of individuals and influence their PA, as discussed, also influences the behavior of those individuals. Thus, individuals' behavior is favorable to entrepreneurship when potential entrepreneurs have the emotional support of those important people in their lives (Ramos-Rodríguez et al., 2019). In this sense, a study in a sample of Italian university students found that the effect of SN on PBC was significant and that it was more robust in the female group than in the male group. (Scafarto et al., 2019). Eyel & Vatansever Durmaz (2019) also found a positive relationship between SN on TPB in undergraduate social sciences students and natural sciences studying at Bahçeşehir University. Moreover, although only partially, a relationship between SN and PBC was also found in a study in a sample of international students at universities throughout Turkey (Usman & Yennita, 2019). This previous evidence justifies to propose the following hypothesis :

**H5:** SN is positively related to PBC.

Regarding the influence of PA on PBC, Usman & Yennita (2019) found a strong effect of PA on PBC in college students. Similarly, in a sample of high school students in Portugal, the relationship of PA on PBC was found to be significant (Do Paço et al., 2011). Furthermore, in a study among farmers that analyzed the factors affecting the intention to continue with Conservation Agriculture during 2020, Tama et al. (2021) found scientific evidence of PA's relationship with PBC. These arguments allow formulating the following research hypothesis:

**H7:** PA toward entrepreneurship is positively related to PBC.

As reflected in the entrepreneurship literature within the TPB framework, PA, SN, and PBC are determinants of entrepreneurial intention. (Krueger et al., 2000; Liñan, 2004; Soomro et al., 2018; Urban & Ratsimanetrimanana, 2019). In this sense, PA is the determinant that motivates intentions in more studies. However, the relationship between SN and intentions is the one that is most often not supported. Of the three factors, the one that usually has the most significant relationship with intentions is PBC, while SN usually has the weakest impact. (Lortie & Castogiovanni, 2015). Nevertheless, we have found sufficient empirical evidence that PA, SN and PBC are antecedents of entrepreneurial intentions. Thus, Kautonen et al. (2011) found a positive relationship of the three factors on entrepreneurial intentions in their study of the working-age population in Finland. Similarly, in an investigation in a sample of students at a university in Yemen, the three predictors of TPB intention were found to be significant (Al-Jubari, 2019). In addition, in the analysis of data obtained through a questionnaire answered by Information System students, it was also found that the three variables that precede intentions in the TPB impact the students' intention to create a new business. (Kaltenecker et al., 2015). Based on the arguments presented, we propose the following relationships:

**H6:** SN is positively related to entrepreneurial intention.

**H8:** PA toward entrepreneurship is positively related to entrepreneurial intention.

**H9:** PBC is positively related to entrepreneurial intention.

Figure 1 shows the relationships considered in the hypotheses we have put forward.

**Figure 1. Research model**

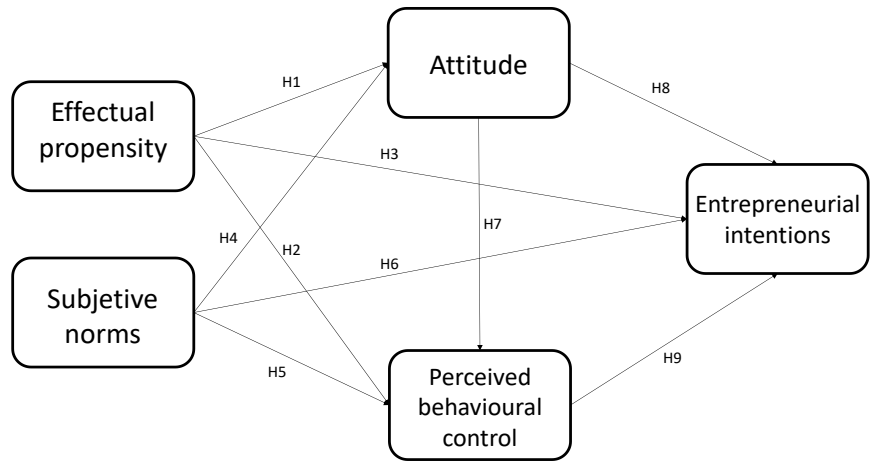

## 3. Methodology

### 3.1. Data collection

The tourism degree students, who voluntarily participated in the research, were interviewed using a self-administered questionnaire using non-probabilistic sampling. Experts and researchers reviewed the questionnaire for content validity and included a pilot test among 30 students. The questions were formulated using a 7-point Likert scale, where one means strongly disagree, and seven means strongly agree. After data collection, we obtained a sample of 464 correctly completed questionnaires.

Of the sample, 56.71% are students from the University of Seville, while 43.29% are from Cadiz. Concerning the sample, 71.02% of the respondents are women compared to 28.98% men. Between 18 and 21 years old are 56.04% of the cases, 37.58% are between 22 and 25 years old, and the rest are older than 25 years old; most of them are Spanish, only 2.6% are foreigners; 31.82% of those surveyed belong to a family whose parents have had or have their own business. Moreover, 34.20% have worked or are currently working, compared to 46.84% who have never worked. The rest have worked, are currently working, or are self-employed or work both for themselves and others.

### 3.2. Measurement scales

According to Sarasvathy (2001) y Gabrielsson & Politis (2011), the effectual orientation is measured independently of the causal orientation. This research focuses on the effectual orientation employing the EP construct. Therefore, we determined the scale with the indicators or items necessary to measure this variable, which is not directly observable by the researcher. The indicators of the EP construct were adapted from the work of Martín-Navarro et al. (2021) and are integrated into five subdimensions: (1) means orientation which is measured through three items, (2) association orientation was measured through four (3) affordable loss orientation is composed of three indicators (4) contingency orientation of four and, (5) control orientation has been measured with four items. To measure the constructs of the Theory of Planned Behavior

(PA, PBC, SN and EI), we used an adapted version of the Entrepreneurial Intention Questionnaire (EIQ) (Liñán et al., 2011). Thus, the four variables belonging to the TPB were measured with the following number of items: (1) PA with five indicators, (2) PBC likewise with five items, (3) SN was measured with four indicators and, (4) EI with six. As proposed by the authors, to avoid the problems of response set bias and the halo effect, the items were randomly mixed and ordered in the questionnaire.

## 4. RESULTS
### 4.1. Analysis of the measurement model

PLS-SEM allows the estimation of complex models with many constructs, indicator variables, and structural trajectories without imposing distributional assumptions on the data (Roldán & Sánchez-Franco, 2012). PLS is a recommended method for studying latent construct models made up of composite (Rigdon, 2016). For the model analysis, we have used the SmartPLS 3 (Ringle et al., 2015) software, which allows us to analyze the relationships between latent variables and their indicators, i.e., how these constructs measure their respective indicators. Likewise, PLS evaluates the measurement of the variables based on individual reliability, construct reliability, discriminant validity and convergent validity. Reliability ensures that the measurement produces consistent results, and validity ensures that the indicators of a construct measure the construct they are intended to measure and not another.

The reliability of individual items must be satisfied that the loadings are greater than 0.707 (Carmines & Zeller, 1979). As Table 1 shows, only the means orientation (MO) in the EP has a lower loading and will be eliminated from the analysis. Subsequently, the reliability of the construct is analyzed through Dijkstra-Henseler rho_A in all cases with values higher than 0.7. (Dijkstra & Henseler, 2015). The composite reliability (Werts et al., 1974) should be higher than 0.8, as Nunnally & Bernstein (1995) suggested. To complete this analysis, convergent validity is checked through the Average Variance Extracted (AVE). In this case, the values must be greater than 0.5 (Fornell & Larcker, 1981), and this is the case.

**Table 1. Reliability and convergent validity**

| Construct/Indicator | Loads | rho_A | Compound reliability | AVE |
|---|---|---|---|---|
| **Effectual propensity (EP)** | | 0.766 | 0.831 | 0.554 |
| Media orientation | 0.642 | | | |
| Guidance to the association | 0.742 | | | |
| Contingency orientation | 0.821 | | | |
| Control orientation | 0.760 | | | |
| **Personal Attitude (PA)** | | 0.922 | 0.938 | 0.751 |
| Being an entrepreneur has more advantages than disadvantages for me. | 0.815 | | | |
| A career in business is attractive to me | 0.906 | | | |
| If I had the opportunity and the resources, I would like to start a company. | 0.856 | | | |
| Being an entrepreneur would bring me great satisfaction. | 0.863 | | | |
| Among several options, I would prefer to be an entrepreneur | 0.891 | | | |
| **Perceived Behavioral Control (PBC)** | | 0.912 | 0.935 | 0.741 |
| Starting a business and keeping it running would be easy for me. | 0.820 | | | |
| I am ready to create a viable business | 0.897 | | | |
| I can control the process of setting up a new company. | 0.895 | | | |
| I know how to develop a business project | 0.861 | | | |
| If I tried to start a company, I would have a good chance of success. | 0.828 | | | |
| **Subjective norms (SN)** | | 0.872 | 0.921 | 0.796 |
| If you decided to start a business, would your immediate family approve of that decision? | 0.879 | | | |
| If you decided to start a company, would your friends approve of that decision? | 0.918 | | | |
| If you decided to start a company, would your peers approve of that decision? | 0.879 | | | |

| Entrepreneurial Intention (EI) | | 0.958 | 0.966 | 0.827 |
|---|---|---|---|---|
| I am willing to do anything to be an entrepreneur. | 0.851 | | | |
| My career goal is to become an entrepreneur | 0.931 | | | |
| I will do my best to start and run my own company. | 0.918 | | | |
| I am determined to start a company in the future | 0.936 | | | |
| I have been thinking very seriously about starting a company | 0.903 | | | |
| I have the firm intention of creating a company someday. | 0.913 | | | |

Table 2 shows the discriminant validity analysis using the heterotraitomonotrait (HTMT) relationship. (Henseler et al., 2016). This ratio must have values below 0.85 (Gold et al., 2001) to ensure that discriminant validity is satisfied, as can be verified.

**Table 2. Discriminant validity (HTMT)**

|  | EN | PC | EO | EN | SN |
|---|---|---|---|---|---|
| EN | | | | | |
| PBC | 0,560 | | | | |
| EP | 0,448 | 0,407 | | | |
| EI | 0,805 | 0,641 | 0,441 | | |
| SN | 0,415 | 0,332 | 0,444 | 0,370 | |

In the second-order construct, the EP, in the first step, eliminated the dimension "affordable loss orientation" because the loads are lower than 0.707. In this second step, the dimension "means orientation" is eliminated for the same reason. Once the measurement model has been refined, we continue with the analysis of the structural model.

### 4.2. Structural model

After analyzing the reliability and validity, the structural model should be evaluated to identify the relationships between the constructs of the proposed research model (Velicia Martín et al., 2020). This analysis tests whether the hypotheses and the relationships of the constructs of the theoretical model are supported (Henseler et al., 2015). The evaluation of the structural model involves the estimation of the path loadings and the values of $R^2$. For this purpose, a bootstrapping technique with 5,000 resamples is used to test the proposed hypotheses (Hair et al., 2011).

The path loadings indicate the strength of the relationships between the explanatory variables and the variables to be explained (Ramírez-Correa et al., 2019). The results of the path loadings are shown in Table 3. In this sense, it is found that out of the nine relationships proposed in the theoretical model, seven of them are significant. The unsupported hypotheses are H5 and H6. Therefore, the relationship SN on PBC is not significant, nor SN on EI.

**Table 3. Results of significance tests of the coefficients of the structural model.**

| Hypothesis | β (Standard Path Coeff.) | T Statistics | P Values | CI | Sig | |
|---|---|---|---|---|---|---|
| H1: EP→AT | 0,279 | 6,489 | 0,000 | (0,198;0,365) | Yes | *** |
| H2: EP→PBC | 0,175 | 3,90 | 0,000 | (0,085;0,262) | Yes | *** |
| H3: EP→IN | 0,077 | 2,407 | 0,016 | (0,015;0,138) | Yes | ** |
| H4: SN→AT | 0,277 | 5,297 | 0,000 | (0,171;0,379) | Yes | *** |
| H5: SN →PBC | 0,084 | 1,861 | 0,063 | (-0,006;0,171) | No | n.s. |
| H6: SN →EI | 0,016 | 0,534 | 0,593 | (-0,043;0,075) | No | n.s. |
| H7: AT→PBC | 0,425 | 10,345 | 0,000 | (0,344;0,504) | Yes | *** |
| H8: AT→EI | 0,587 | 18,552 | 0,000 | (0,523;0,647) | Yes | *** |
| H9: PBC→ EI | 0,264 | 7,382 | 0,000 | (0,193;0,333) | Yes | *** |

Note: Significant at p***=<0.001, p**<0.05

$R^2$ values measure the predictive power of structural models. Interpreted as multiple regression results, the $R^2$ indicates the variance explained by the exogenous variables (Ramírez-Correa et al., 2019). $R^2$ values are shown in Table 4. Hair et al. (2014) $R^2$ values above 0.67 are considered high, between 0.67 and 0.33 moderate, between 0.33 and 0.19 weak and values below 0.19 are unacceptable. In our case, Entrepreneurial Intention presents a moderate $R^2$, while the values for PA and PBC are weak.

**Table 4. $R^2$-values**

|  | $R^2$ | $R^2$ adjusted |
|---|---|---|
| Personal Attitude | 0.206 | 0.202 |
| Perceived behavioral control | 0.310 | 0.305 |
| Entrepreneurial intention | 0.640 | 0.637 |

The global model fit is performed using the values of the standardized root mean square residual (SRMR). A value of 0 would represent a perfect fit, although values below 0.08 present a good fit (Henseler et al., 2014). In our model, SRMR presents a value of 0.052 lower than the reference value, so our model presents a good fit.

As for indirect effects, there are some significant ones among the model constructs (see Table 5). In addition to showing a direct and positive relationship with EI, EP shows significant indirect effects through PA and PBC. Despite not showing a significant direct relationship with EI or PBC, SN shows a significant indirect relationship through PA with EI and PBC variables.

**Table 5. Indirect effects**

| Specific indirect effects | Route (β) | Statistics T | p-value | CI |
|---|---|---|---|---|
| **EP -> AT -> IN** | 0.164 | 6.204 | 0.000 | (0.110.0.216) |
| **EP -> PBC -> IN** | 0.046 | 3.342 | 0.001 | (0.022.0.077) |
| **EP -> AT -> PBC -> IN** | 0.031 | 4.506 | 0.000 | (0.019.0.046) |
| **EP -> AT -> PBC** | 0.118 | 6.051 | 0.000 | (0.081.0.157) |
| **AT -> PBC -> IN** | 0.112 | 5.817 | 0.000 | (0.077.0.152) |
| **SN -> AT -> IN** | 0.162 | 4.973 | 0.000 | (0.100.0.228) |
| **SN -> PBC -> IN** | 0.022 | 1.846 | 0.065 | (0.000.0.048) |
| **SN -> AT -> PBC -> IN** | 0.031 | 3.759 | 0.000 | (0.017.0.050) |
| **SN -> AT -> PBC** | 0.118 | 4.475 | 0.000 | (0.071.0.174) |

Importance-performance matrix analysis (IPMA) is carried out (see Table 6 and Figure 2). The IPMA assists the PLS-SEM results through a four-quadrant diagram depicted in Figure2. The vertical axis represents attribute performance, from poor performance to good performance. The horizontal axis represents the perceived importance of the attributes from unimportant to very important. (García-Fernández et al., 2020).

**Table 6. IPMA (Business Intention)**

|  | Total effect (importance) | Index value (yield) |
|---|---|---|
| Effectual propensity | 0.711 | 63.127 |
| Personal Attitude | 0.859 | 66.411 |
| Perceived behavioral control | 0.331 | 45.520 |
| Social norm | 0.312 | 75.353 |
| Mean | 0.553 | 62.603 |

The quadrants are delimited using the mean performance and mean importance listed in the PEI results table. As shown in figure 2, the third quadrant is SN, the variable with the highest importance, but it contributes the least to the EI effect, as does PBC, which is in the fourth quadrant. The variable with the most significant contribution to the total effect is PA, followed by EP.

**Figure 2. IPMA (Business Intention)**

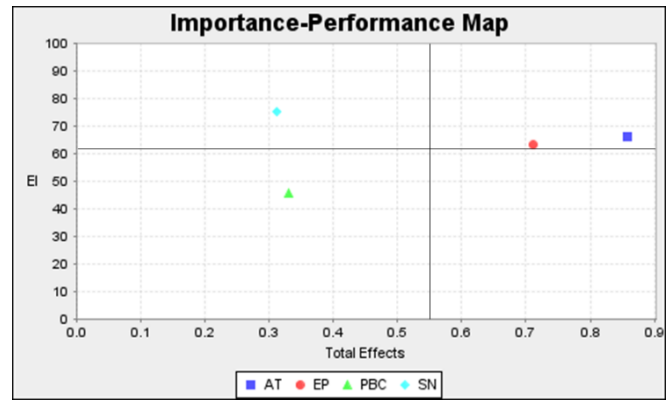

Finally, the model's predictive power is analyzed, which allows testing the model's generalizability to other populations. We apply a PLSpredict, an out-of-sample approach developed. (Shmueli et al., 2016). Following Roldan & Cepeda (2017) with the PLSpredictive technique, we obtained results expressed through the variables RMSE (R Mean Square Error) and MAE (Mean Absolute Error) and Q_predict$^2$. For the model to have predictive power, the value of $Q^2$ must be greater than zero. In our case, all the values in the last column are positive (see table 7). Similarly, we obtained positive values for RMSE and MAE, showing good predictability. (Woodside, 2013). Therefore, this research demonstrated the predictive validity, or out-of-sample predictability, of the proposed model to predict the values of new cases.

**Table 7. Evaluation of PLS predictions**

| | RMSE | MAE | Q_prediction$^2$ |
|---|---|---|---|
| Attitude | 0.901 | 0.710 | 0.196 |
| Perceived behavioral control | 0.922 | 0.757 | 0.156 |
| Entrepreneurial intention | 0.901 | 0.749 | 0.195 |

## 5. DISCUSSION

The academic literature has studied effectuation logic during the entrepreneurial process. That is, effectuation theory has analyzed the behavior of entrepreneurs who have already started their businesses. However, this behavior has not been analyzed in people who have not started a business and are in an early stage. This research integrates the entrepreneurial intention model with the effectual logic. For this purpose, an extended TPB model with the Effectual Propensity variable has been developed. The extended TPB model with EP can be considered a valuable model for predicting entrepreneurial intention in tourism. As shown, the theoretical model has satisfactory psychometric qualities. In addition, the TPB model has increased its predictive power with the EP variable.

In this paper, it was found in a sample of university students that PE is a determinant of EI, both directly and indirectly through PA and PBC. Similarly, it is confirmed that both PA and PBC present a direct and positive relationship with entrepreneurial intention in the sample analyzed. These findings are in line with the results of many other authors in the area of entrepreneurship (Krueger et al., 2000; Liñan, 2004; Soomro et al., 2018; Urban & Ratsimanetrimanana, 2019). Furthermore, even with those of Ahamed & Limbu (2018) in their study of the intention to use credit cards. PE presents a positive and direct relationship with EI and significant indirect effects with the variables PA and PBC that contribute to explaining entrepreneurial intention. The direct relationship between SN and EI shows a non-significant relationship. The same occurs in other studies, such as that of Doanh & Bernat (2019), which did not find the effect of SN on EI among university students in Vietnam. However, it is in contrast to other studies that have found this relationship significant in students' entrepreneurial intentions. (Al-Jubari, 2019; Ridha & Wahyu, 2017). On the other hand, the indirect effect of SN through PA is significant on both EI and PBC.

It should also be noted that the IPMA analysis shows that attitude is the exogenous variable that contributes most to the total effect of the endogenous variable EI, followed by EP. In contrast, the variable that contributes least to the total effect is PBC. The theoretical model analyzed with a sample of students of the Degree in Tourism presents adequate predictive power. The results obtained in our analysis can be generalized to other samples.

## 6. CONCLUSIONS

There is great interest in understanding the antecedents of the intention to create new companies among academics who study entrepreneurship. Among others, the TPB model stands out as one of the most used by researchers to achieve this goal. In this sense, our research has carried out empirical work. The effectual propensity is proposed as an antecedent of entrepreneurial intentions, together with the rest of the TPB determinants (PA, SN and PBC). Our study has a great added value since we have not found in the literature any study that analyzes EP intentions in people who have not yet become entrepreneurs. The results of the hypothesis testing show two critical findings. First, and as the most important finding, this study has shown that EP influences EI. Moreover second, PE also affects EI indirectly through PA and PBC.

Overall, our work contributes to investigating entrepreneurial intentions by providing a new determinant not previously investigated on EI. Thus, we have shown that EI affects students' intentions towards creating a new venture. Similarly, our work has social implications. Our results differentiate between higher education students' effectual and causal orientation. It will allow adjusting the curricula to contain more personalized training in entrepreneurship and strengthen the effectual orientation of students who require it. In this way, public policies must encourage entrepreneurs in the initial stages of the entrepreneurial process. From the business point of view, our research allows us to determine the orientation shown by professionals in the business world to evaluate the PE of their employees, managers, and even business partners. Both those with whom entrepreneurs already have a relationship and those who require a prior evaluation of their entrepreneurial orientation before assuming future commitments. In this same sense, how the university emits knowledge to the market is through hiring its students. For this reason, it is interesting to differentiate between the orientations of these individuals with entrepreneurial intentions. Thus, the labor market will differentiate, according to its needs, the competencies, behaviors or orientations of the graduates it hires.

This research should be interpreted with some limitations in mind. Because the findings have been the results of a single study, and with students from two geographically close universities, they should not be generalized. It would be advisable to repeat this study in different samples with different educational and cultural contexts and socio-demographic profiles. Nevertheless, these limitations present opportunities for future lines of research. For example, it would be interesting to test these scales in other samples from other countries or cultures. Furthermore, even to consider PE as a moderating variable in other models of entrepreneurial intentions.

## 7. REFERENCES

Ahamed, A. F. M. J., & Limbu, Y. B. (2018). Dimensions of materialism and credit card usage: an application and extension of the theory of planned behavior in Bangladesh. *Journal of Financial Services Marketing*, *23*(3–4), 200–209. https://doi.org/10.1057/s41264-018-0058-5

Ahmad, N. H., Ramayah, T., Mahmud, I., Musa, M., & Anika, J. J. (2019). Entrepreneurship as a preferred career option: Modelling tourism students' entrepreneurial intention. *Education and Training*, *61*(9), 1151–1169. https://doi.org/10.1108/ET-12-2018-0269

Ajzen, I. (2002). Constructing a TpB questionnaire: Conceptual and methodological considerations, 1–14. Retrieved from file:///Users/Riccardo/Documents/Riccardo's Library/Library.papers3/Reports/2002/Ajzen/2002 Ajzen.pdf%5Cnpapers3://publication/uuid/B959DE76-CAA4-474C-996C-D014EB9357D9

Ajzen, I., Netemeyer, R., Ryn, M. Van, & Ajzen, I. (1991). The theory of planned behavior. *Organisational Behavior and Human Decision Processes*, *50*(2), 179–211. https://doi.org/10.1016/0749-5978(91)90020-T

Al-Jubari, I. (2019). College students' entrepreneurial intention: Testing an integrated model of SDT and TPB. *SAGE Open*, *9*(2), 1–15. https://doi.org/10.1177/2158244019853467

Al Mamun, A., Binti Che Nawi, N., Dewiendren, A. A., & Fazira Binti Shamsudin, S. F. (2016). Examining the Effects of Entrepreneurial Competencies on Students' Entrepreneurial Intention. *Mediterranean Journal of Social Sciences*, *7*(2), 119–127. https://doi.org/10.5901/mjss.2016.v7n2p119

Annisa, D. N., Tentama, F., & Bashori, K. (2021). The role of family support and internal locus of control in entrepreneurial intention of vocational high school students. *International Journal of Evaluation and Research in Education*, *10*(2), 381–388. https://doi.org/10.11591/ijere.v10i2.20934

Bird, B. (1988). Implementing Entrepreneurial Ideas: The Case for Intention. *Academy of Management Review*, *13*(3), 442–453. https://doi.org/10.5465/AMR.1988.4306970

Brettel, M., Mauer, R., Engelen, A., & Küpper, D. (2012). Corporate effectuation: Entrepreneurial action and its impact on R&D project performance. *Journal of Business Venturing*, *27*(2), 167–184. https://doi.org/10.1016/j.jbusvent.2011.01.001

Breznitz, S. M., & Zhang, Q. (2022). Entrepreneurship education and firm creation. *Regional Studies*, *56*(6), 940–955. https://doi.org/10.1080/00343404.2021.1878127

Carmines, E. G., & Zeller, R. A. (1979). *Reliability and validity assessment*. Beverly Hills: Sage Publications.

Cava Jimenez, J. A., Millán Vázquez de la Torre, M. G., & Dancausa Millán, M. G. (2022). Enotourism in Southern Spain: The Montilla-Moriles PDO. *International Journal of Environmental Research and Public Health*, *19*(6). https://doi.org/10.3390/ijerph19063393

Cepeda, G., Henseler, J., Ringle, C. M., & Luis, J. (2017). Prediction-oriented modeling in business research by means of PLS path modeling : Introduction to a JBR special section. *Journal of Business Research*, *69*(2016), 4545–4551. https://doi.org/10.1016/j.jbusres.2016.03.048

Chandler, G. N., DeTienne, D. R., McKelvie, A., & Mumford, T. V. (2011). Causation and effectuation

processes: A validation study. *Journal of Business Venturing*, *26*(3), 375–390. https://doi.org/10.1016/j.jbusvent.2009.10.006

Chin, W. W. (1998). Issues and Opinion on Structural Equation Modeling. *MIS Quarterly*, *22*(1), VII–XVI. https://doi.org/Editorial

Chin, W. W., & Newsted, P. R. (1999). Structural equation modeling analysis with small samples using partial least squares. *Statistical Strategies for Small Sample Research*, *1*(1), 307–341.

Chin, W. W., & Todd, P. A. (1995). On the use, usefulness, and ease of use of structural equation modeling in mis research: A note of caution. *MIS Quarterly: Management Information Systems*, *19*(2), 237–246. https://doi.org/10.2307/249690

Dew, N., Read, S., Sarasvathy, S. D., & Wiltbank, R. (2009). Effectual versus predictive logics in entrepreneurial decision-making : Differences between experts and novices. *Journal of Business Venturing*, *24*(4), 287–309. https://doi.org/10.1016/j.jbusvent.2008.02.002

Dijkstra, T. K., & Henseler, J. (2015). Consistent Partial Least Squares Path Modeling. *MIS Quarterly*, *39*(2), 297–316.

Do Paço, A., Ferreira, J., Raposo, M., Rodrigues, R. G., & Dinis, A. (2011). Entrepreneurial intention among secondary students: Findings from Portugal. *International Journal of Entrepreneurship and Small Business*, *13*(1), 92–106. https://doi.org/10.1504/IJESB.2011.040418

Doanh, D. C., & Bernat, T. (2019). Entrepreneurial self-efficacy and intention among vietnamese Entrepreneurial self-efficacy and intention among vietnamese students : a meta-analytic path analysis based on the theory of planned behavior. *Procedia Computer Science*, *159*, 2447–2460. https://doi.org/10.1016/j.procs.2019.09.420

Erickson, S. M., & Laing, W. (2016). The oxford MBA: A case study in connecting academia with business. *Journal of Entrepreneurship Education*, *19*(1), 1–8.

Esfandiar, K., Sharifi-Tehrani, M., Pratt, S., & Altinay, L. (2019). Understanding entrepreneurial intentions: A developed integrated structural model approach. *Journal of Business Research*, *94*(October 2017), 172–182. https://doi.org/10.1016/j.jbusres.2017.10.045

Eyel, C. S., & Vatansever Durmaz, İ. B. (2019). Entrepreneurial Intentions of Generation-Z: Compare of Social Sciences and Natural Sciences Undergraduate Students at Bahçeşehir University. *Procedia Computer Science*, *158*, 861–868. https://doi.org/10.1016/j.procs.2019.09.124

Fishbein, M., & Ajzen, I. (1977). Belief, attitude, intention, and behavior: An introduction to theory and research. *Philosophy and Rhetoric*, *10*(2).

Fong, V. H. I., Wong, I. K. A., & Hong, J. F. L. (2018). Developing institutional logics in the tourism industry through coopetition. *Tourism Management*, *66*, 244–262. https://doi.org/10.1016/j.tourman.2017.12.005

Fornell, C., & Larcker, D. F. (1981). Evaluating Structural Equation Models with Unobservable Variables and Measurement Error. *Journal of Marketing Research*, *18*(1), 39. https://doi.org/10.2307/3151312

Gabrielsson, J., & Politis, Æ. D. (2011). Career motives and entrepreneurial decision-making : examining preferences for causal and effectual logics in the early stage of new ventures. *Small Business Economics*, *36*(3), 281–298. https://doi.org/10.1007/s11187-009-9217-3

García-Fernández, J., Fernández-Gavira, J., Sánchez-Oliver, A. J., Gálvez-Ruíz, P., Grimaldi-Puyana, M., & Cepeda-Carrión, G. (2020). Importance-performance matrix analysis (Ipma) to evaluate servicescape fitness consumer by gender and age. *International Journal of Environmental Research and Public Health*, *17*(18), 1–19. https://doi.org/10.3390/ijerph17186562

García-Machado, J. J., Martín, E. B., & Rengel, C. G. (2020). A PLS multigroup analysis of the role of businesswomen in the tourism sector in Andalusia. *Forum Scientiae Oeconomia*, *8*(2), 37–57. https://doi.org/10.23762/fso

Geisser, S. (1974). A Predictive Approach to the Random Effect Model. *Biometrika*, *61*(1), 101–107.

Gielnik, M. M., & Frese, M. (2013). Entrepreneurship and poverty reduction: Applying I-O psychology to microbusiness and entrepreneurship in developing countries. In *Using Industrial-Organisational Psychology for the Greater Good: Helping those who Help others* (pp. 394–438). New York, NY, USA: Routledge, Taylor and Francis,. https://doi.org/10.4324/9780203069264

Gold, A. A. H., Malhotra, A., & Segars, A. A. H. (2001). Knowledge management: An organisational capabilities perspective. *Journal of Management Information Systems*, *18*(1), 185–214. https://doi.org/10.1002/ceat.201000522

Guo, R., Cai, L., & Zhang, W. (2016). Effectuation and causation in new internet venture growth: The mediating effect of resource bundling strategy. *Internet and Higher Education*, *26*(2), 460–483. https://doi.org/10.1108/IntR-01-2015-0003

Hair, J. F., Risher, J. J., Sarstedt, M., & Ringle, C. M. (2019). When to use and how to report the results of PLS-SEM. *European Business Review*, *31*(1), 2–24. https://doi.org/10.1108/EBR-11-2018-0203

Hair Jr, J. F. (2021). Next-generation prediction metrics for composite-based PLS-SEM. *Industrial Management and Data Systems*, *121*(1), 5–11. https://doi.org/10.1108/IMDS-08-2020-0505

Hair Jr, J. F., Black, W. C., Babin, B. J., & Anderson, R. E. (2014). *Multivariate data analysis (Pearson new internat. ed)*. New Jersey: Essex: Pearson.

Hair Jr, J. F., Ringle, C. M., & Sarstedt, M. (2011). PLS-SEM : Indeed a Silver Bullet. *Journal of Marketing Theory and Practice*, *19*(2), 139–151. https://doi.org/10.2753/MTP1069-6679190202

Hair Jr, J. F., Sarstedt, M., Hopkins, L., & Kuppelwieser, V. G. (2014). Partial least squares structural equation modeling (PLS-SEM). *European Business Review*, *26*(2), 106–121. https://doi.org/10.1108/EBR-10-2013-0128

Henseler, J., Dijkstra, T. K., Sarstedt, M., Ringle, C. M., Diamantopoulos, A., Straub, D. W., … Calantone, R. J. (2014). Common Beliefs and Reality About PLS: Comments on Ronkko and Evermann (2013). *Organizational Research Methods*, *17*(2), 182–209. https://doi.org/10.1177/1094428114526928

Henseler, J., Ringle, C. M., & Sarstedt, M. (2014). A new criterion for assessing discriminant validity in variance-based structural equation modeling. *Journal of the Academy of Marketing Science*, *43*(1), 115–135. https://doi.org/10.1007/s11747-014-0403-8

Henseler, J., Ringle, C. M., & Sarstedt, M. (2015). A new criterion for assessing discriminant validity in variance-based structural equation modeling. *Journal of the Academy of Marketing Science*, *43*(1), 115–135. https://doi.org/10.1007/s11747-014-0403-8

Henseler, J., Ringle, C. M., & Sarstedt, M. (2016). Testing measurement invariance of composites using partial least squares. *International Marketing Review*, *33*(3), 405–431. https://doi.org/10.1108/IMR-09-2014-0304

Henseler, J., Ringle, C. M., & Sinkovics, R. (2009). THE USE OF PARTIAL LEAST SQUARES PATH MODELING IN INTERNATIONAL MARKETING. *Advances in International Marketing*, *20*, 277–319. https://doi.org/10.1016/0167-8116(92)90003-4

Houssain Bouichou, E., Abdoulaye, T., Allali, K., Bouayad, A., & Fadlaoui, A. (2021). Entrepreneurial Intention among Rural Youth in Moroccan Agricultural Cooperatives: The Future of Rural Entrepreneurship. *Sustainability*, *13*(6). https://doi.org/10.3390/su13169247

Joreskog, K. G., & Wold, H. O. A. (1982). The ML and PLS techniques for modeling with latent variables: Historical and comparative aspects. In *Systems under indirect observation, part I* (pp. 263–270). North-Holland.

Kaltenecker, N., Hoerndlein, C., & Hess, T. (2015). The drivers of entrepreneurial intentions - an empirical study among information systems and computer science students. *Journal of Entrepreneurship Education*, *18*(2), 39–52.

Kautonen, T., van Gelderen, M., & Tornikoski, E. T. (2013). Predicting entrepreneurial behaviour: a test of the theory of planned behaviour. *Applied Economics*, *45*(August 2014), 697–707. https://doi.org/10.1080/00036846.2011.610750

Kitagawa, F., Marzocchi, C., Sánchez-Barrioluengo, M., & Uyarra, E. (2022). Anchoring talent to regions: the role of universities in graduate retention through employment and entrepreneurship. *Regional Studies*, *56*(6), 1001–1014. https://doi.org/10.1080/00343404.2021.1904136

Kotler, P., & Keller, K. L. (2015). *Marketing Management eBook*. Pearson Higher Ed.

Krueger, N. F. (2007). What Lies Beneath ? The Experiential Essence of Entrepreneurial Thinking. *Entrepreneurship Theory and Practice*, *31*(1), 123–138.

Krueger, N. F., Reilly, M. D., & Carsrud, A. L. (2000). Competing models of entrepreneurial intentions. *Journal of Business Venturing*, *15*(5–6), 411–432. https://doi.org/10.1016/S0883-9026(98)00033-0

Kusumawardani, K. A., Widyanto, H. A., & Deva, I. P. L. I. (2020). Understanding the Entrepreneurial Intention of Female Entrepreneurs in the Balinese Tourism Industry. *International Journal of Research in Business and Social Science (2147- 4478)*, *9*(1), 63–79. https://doi.org/10.20525/ijrbs.v9i1.611

Lateh, M., Hussain, M. D., & Halim, M. S. A. (2017). Micro Enterprise Development and Income Sustainability for Poverty Reduction: a Literature Investigation. *Journal of Business and Technopreneurship*, *7*(1), 23–38.

Lechuga Sancho, M. P., Martín-Navarro, A., & Ramos-Rodríguez, A. R. (2020). Information Systems Management Tools: An Application of Bibliometrics to CSR in the Tourism Sector. *Sustainability*, *12*(20), 8697. https://doi.org/10.3390/su12208697

Lechuga Sancho, M. P., Ramos-Rodríguez, A. R., & Frende Vega, M. Á. (2021). Is a favorable entrepreneurial climate enough to become an entrepreneurial university? An international study with GUESSS data. *International Journal of Management Education*, *19*(3), 100536. https://doi.org/10.1016/j.ijme.2021.100536

Li, L., & Wu, D. (2019). Entrepreneurial education and students' entrepreneurial intention: does team cooperation matter? *Journal of Global Entrepreneurship Research*, *9*(1–13).

Liñan, F. (2004). Intention-based models of entrepreneurship education. *Piccola Impresa/ Small Business*, *3*(January 2004), 11–35.

Liñán, F. (2008). Skill and value perceptions: How do they affect entrepreneurial intentions? *International Entrepreneurship and Management Journal*, *4*(3), 257–272. https://doi.org/10.1007/s11365-008-0093-0

Liñán, F., & Chen, Y. (2006). Testing the Entrepreneurial Intention Model on a two-country Sample. *Documents de Treball*, *06/7*, 1–37.

Liñán, F., & Fayolle, A. (2015). A systematic literature review on entrepreneurial intentions: citation, thematic analyses, and research agenda. *International Entrepreneurship and Management Journal*, 907–933. https://doi.org/10.1007/s11365-015-0356-5

Liñán, F., Rodríguez-Cohard, J. C., & Rueda-Cantuche, J. M. (2011). Factors affecting entrepreneurial

intention levels: A role for education. *International Entrepreneurship and Management Journal*, *7*(2), 195–218. https://doi.org/10.1007/s11365-010-0154-z

Lohmöller, J.-B. (1989). Predictive vs. Structural Modeling: PLS vs. ML. *Latent Variable Path Modeling with Partial Least Squares*, (1983), 199–226. https://doi.org/10.1007/978-3-642-52512-4_5

Lortie, J., & Castogiovanni, G. (2015). The theory of planned behavior in entrepreneurship research: what we know and future directions. *International Entrepreneurship and Management Journal*, 935–957. https://doi.org/10.1007/s11365-015-0358-3

Maheshwari, G. (2021). Factors influencing entrepreneurial intentions the most for university students in Vietnam: educational support, personality traits or TPB components? *Education and Training*, *63*(7–8), 1138–1153. https://doi.org/10.1108/ET-02-2021-0074

Martín-Navarro, A., Medina-Garrido, J. A., & Velicia-Martín, F. (2021). How effectual will you be? Development and validation of a scale in higher education. *International Journal of Management Education*, *19*(3), 100547. https://doi.org/10.1016/j.ijme.2021.100547

Matalamäki, M., Vuorinen, T., Varamäki, E., & Sorama, K. (2017). Business Growth in Established Companies; Roles of Effectuation and Causation. *Journal of Enterprising Culture*, *25*(02), 123–148. https://doi.org/10.1142/S0218495817500054

Mei, H., Zhan, Z., Fong, P. S. W., Liang, T., & Ma, Z. (2016). Planned behaviour of tourism students' entrepreneurial intentions in China. *Applied Economics*, *48*(13), 1240–1254. https://doi.org/10.1080/00036846.2015.1096006

Mueller, S. L., & Thomas, A. S. (2001). Culture and entrepreneurial potential: A nine country study of locus of control and innovativeness. *Journal of Business Venturing*, *16*(1), 51–75. https://doi.org/10.1016/S0883-9026(99)00039-7

Nasip, S., Amirul, S. R., Sondoh, S. L., & Tanakinjal, G. H. (2017). Psychological characteristics and entrepreneurial intention: A study among university students in North Borneo, Malaysia. *Education and Training*, *59*(7–8), 825–840. https://doi.org/10.1108/ET-10-2015-0092

Nunnally, J. C., & Bernstein, I. J. (1995). *Teoría Psicométrica (3ª ed)*. México, D.F.: McGraw-Hill Latinomericana.

Perry, J. T., Chandler, G. N., & Markova, G. (2012). Entrepreneurial Effectuation: A Review and Suggestions for Future Research. *Entrepreneurship Theory and Practice*, *36*(4), 837–861. https://doi.org/10.1111/j.1540-6520.2010.00435.x

Phuc, P. T., Vinh, N. Q., & Do, Q. H. (2020). Factors affecting entrepreneurial intention among tourism undergraduate students in Vietnam. *Management Science Letters*, *10*(15), 3675–3682. https://doi.org/10.5267/j.msl.2020.6.026

Piroth, P., Ritter, M. S., & Rueger-Muck, E. (2020). Online grocery shopping adoption: do personality traits matter? *British Food Journal*, *122*(3), 957–975. https://doi.org/10.1108/BFJ-08-2019-0631

Qureshi, M. S., & Mahdi, F. (2014). Impact of Effectuation Based Interventions on the Intentions to Start a Business. *IBA Business Review*, *9*(2), 143–157.

Ramírez-Correa, P., Rondán-Cataluña, F. J., Arenas-Gaitán, J., & Martín-Velicia, F. (2019). Analysing the acceptation of online games in mobile devices: An application of UTAUT2. *Journal of Retailing and Consumer Services*, *50*(May), 85–93. https://doi.org/10.1016/j.jretconser.2019.04.018

Ramos-Rodríguez, A. R., Medina-Garrido, J. A., & Ruiz-Navarro, J. (2019). Why not now? Intended timing in entrepreneurial intentions. *International Entrepreneurship and Management Journal*,

*15*(4), 1221–1246. https://doi.org/10.1007/s11365-019-00586-5

Read, S., & Sarasvathy, S. D. (2005). Knowing What to Do and Doing What You Know. *The Journal of Private Equity*, *9*(1), 45–62. https://doi.org/10.3905/jpe.2005.605370

Reyes-Menendez, A., Palos-Sanchez, P. R., Saura, J. R., & Martin-Velicia, F. (2018). Understanding the Influence of Wireless Communications and Wi-Fi Access on Customer Loyalty: A Behavioral Model System. *Wireless Communications and Mobile Computing*, *2018*. https://doi.org/10.1155/2018/3487398

Ridha, R.N.; Wahyu, B. P. (2017). Entrepreneurship intention in agricultural sector of young generation in Indonesia. *Asia Pacific Journal of Innovation and Entrepreneurship*, *11*(1), 76–89. https://doi.org/10.1108/APJIE-04-2017-022

Rigdon, E. E. (2016). Choosing PLS path modeling as analytical method in European management research : A realist perspective. *European Management Journal*, *34*(6), 598–605. https://doi.org/10.1016/j.emj.2016.05.006

Ringle, C. M., Wende, S., & Becker, J. M. (2015). SmartPLS 3. Boenningstedt: SmartPLS GmbH.

Roldán, J. L., & Sánchez-Franco, M. J. (2012). Variance-based structural equation modeling: Guidelines for using partial least squares in information systems research. In *Research methodologies, innovations and philosophies in software systems engineering and information systems* (pp. 193–221). IGI Global.

Roldan, J.L., Cepeda, G. (2017). *PLS-SEM*. CFP, Universidad de Sevilla, fourth ed.

Sabirov, K. (2022). The role of small business in reducing poverty. *Eurasian Journal of Law, Finance and Applied Sciences*, *2*(2), 269–273. https://doi.org/10.5281/zenodo.6346743

Sanchez Franco, M. J., Martin Velicia, F. A., & Villarejo Ramos, A. (2007). The TAM model and higher learning: a study on the moderating effect of, gender. *Revista Espanola de Pedagogia*, *65*(238), 459–478.

Sarasvathy, S. D. (2001a). Causation and Effectuation: Toward a Theoretical Shift From Economic Inevitability To Entrepreneurial Contingency. *Academy of Management Review*, *26*(2), 243–263. https://doi.org/10.5465/AMR.2001.4378020

Sarasvathy, S. D. (2001b). Effectual Reasoning in Entrepreneurial Decision Making: Existence and Bounds. *Academy of Management Proceedings*, *2001*(1), D1–D6. https://doi.org/10.5465/apbpp.2001.6133065

Sarasvathy, S. D., & Dew, N. (2008). Effectuation and Over-Trust : Debating Goel and Karri. *Entrepreneurship Theory and Practice*, *32*(4), 727–737.

Sarstedt, M., Hair, J. F., Cheah, J. H., Becker, J. M., & Ringle, C. M. (2019). How to specify, estimate, and validate higher-order constructs in PLS-SEM. *Australasian Marketing Journal*, *27*(3), 197–211. https://doi.org/10.1016/j.ausmj.2019.05.003

Scafarto, F., Poggesi, S., & Mari, M. (2019). Entrepreneurial Intentions, Risk-Taking Propensity and Environmental Support: The Italian Experience. In *The Anatomy of Entrepreneurial Decisions. Contributions to Management Science.* (pp. 214–234). https://doi.org/10.1007/978-3-030-19685-1_10

Schmidt, J., & Heidenreich, S. (2014). Investigating Organisational Antecedents of Effectual Corporate Entrepreneurship. In *ISPIM Conference Proceedings; Manchester* (pp. 1–33).

Shahidi, N. (2020). The Moderating Effects of Sustainability Orientation in the Entrepreneurial Intention Model. *Journal of Enterprising Culture*, *28*(01), 59–79. https://doi.org/10.1142/S021849582050003X

Shahzad, M. F., Khan, K. I., Saleem, S., & Rashid, T. (2021). What factors affect the entrepreneurial intention to start-ups? The role of entrepreneurial skills, propensity to take risks, and innovativeness in open business models. *Journal of Open Innovation: Technology, Market, and Complexity*, *7*(3). https://doi.org/10.3390/JOITMC7030173

Shapero, A. (1984). The entrepreneurial event. In *C. A. Kent (Ed.), The environment for entrepreneurship*. Lexington, Mass.: Lexington Books.

Shmueli, G., Ray, S., Manuel, J., Estrada, V., & Chatla, S. B. (2016). The elephant in the room : Predictive performance of PLS models. *Journal of Business Research*, *69*(10), 4552–4564. https://doi.org/10.1016/j.jbusres.2016.03.049

Smolka, K. M., Verheul, I., Burmeister-Lamp, K., & Heugens, P. P. M. A. R. (2018). Get it together! Synergistic effects of causal and effectual decision-making logics on venture performance. *Entrepreneurship: Theory and Practice*, *42*(4), 571–604. https://doi.org/10.1111/etap.12266

Somi, J. (2015). *The influence of entrepreneurial intentions on causation and effectuation*.

Soomro, B. A., Shah, N., & Memon, M. (2018). Robustness of the Theory of Planned Behavior (TPB): A Comparative study between Pakistan and Thailand. *Academy of Entrepreneurship Journal*, *24*(3), 1–18.

Tama, R. A. Z., Ying, L., Yu, M., Hoque, M. M., Adnan, K. M., & Sarker, S. A. (2021). Assessing farmers' intention towards conservation agriculture by using the Extended Theory of Planned Behavior. *Journal of Environmental Management*, *280*(March 2020). https://doi.org/10.1016/j.jenvman.2020.111654

Thompson, E. R. (2009). Individual entrepreneurial intent: Construct clarification and development of an internationally reliable metric. *Entrepreneurship: Theory and Practice*, *33*(3), 669–695.

Tleuberdinova, A., Shayekina, Z., Salauatova, D., & Pratt, S. (2021). Macro-economic Factors Influencing Tourism Entrepreneurship: The Case of Kazakhstan. *Journal of Entrepreneurship*, *30*(1), 179–209. https://doi.org/10.1177/0971355720981431

Torres, F. C., Méndez, J. C. E., Barreto, K. S., Chavarría, A. P., Machuca, K. J., & Guerrero, J. A. O. (2017). Exploring entrepreneurial intentions in Latin American university students. *International Journal of Psychological Research*, *10*(2), 46–59. https://doi.org/10.21500/20112084.2794

Trang, T. Van, & Doanh, D. C. (2019). The role of structural support in predicting entrepreneurial intention: Insights from Vietnam. *Management Science Letters*, *9*(11), 1783–1798. https://doi.org/10.5267/j.msl.2019.6.012

Tsai, K. H., Chang, H. C., & Peng, C. Y. (2016). Refining the linkage between perceived capability and entrepreneurial intention: roles of perceived opportunity, fear of failure, and gender. *International Entrepreneurship and Management Journal*, *12*(4), 1127–1145. https://doi.org/10.1007/s11365-016-0383-x

Urban, B., & Ratsimanetrimanana, F. (2019). Access to finance and entrepreneurial intention: An empirical study of Madagascan rural areas. *Journal of Enterprising Communities*, *13*(4), 455–471. https://doi.org/10.1108/JEC-12-2018-0106

Usman, B., & Yennita. (2019). Understanding the entrepreneurial intention among international students in. *Journal of Global Entrepreneurship Research*, *9*(1), 1–21.

Valliere, D. (2014). Purifying entrepreneurial intent: proposition of a new scale. *Journal of Small Business and Entrepreneurship*, *27*(5), 451–470. https://doi.org/10.1080/08276331.2015.1094225

Vargas-Hernández, J. G., Pérez, C. O. M., & Íñiguez, M. A. E. (2021). Green innovation business (GIB)

as a comprehensive entrepreneurship model for internationalisation. The case study of BIO-FOM in the urban area of Guadalajara. *International Journal of Trade & Commerce-IIARTC*, *10*(1), 1–16. https://doi.org/10.46333/ijtc/10/1/1

Velicia Martín, F., Toledo, L. D., & Palos-Sanchez, P. (2020). How deep is your love? Brand love analysis applied to football teams. *International Journal of Sports Marketing and Sponsorship*, *21*(4), 669–693. https://doi.org/10.1108/IJSMS-10-2019-0112

Villani, E., Linder, C., & Grimaldi, R. (2018). Effectuation and causation in science-based new venture creation: A configurational approach. *Journal of Business Research*, *83*(October 2017), 173–185. https://doi.org/10.1016/j.jbusres.2017.10.041

Werts, C. E., Linn, R. L., & Jöreskog, K. G. (1974). Intraclass reliability estimates: Testing structural assumptions. *Educational and Psychological Measurement*, *34*(1), 25–33. https://doi.org/10.1177/001316447403400104

Woodside, A. G. (2013). Moving beyond multiple regression analysis to algorithms: Calling for adoption of a paradigm shift from symmetric to asymmetric thinking in data analysis and crafting theory. *Journal of Business Research*, *66*(4), 463–472. https://doi.org/10.1016/j.jbusres.2012.12.021

Yasa, N. N. K., Rahmayanti, P. L. D., Telagawathi, N. L. W. S., Witarsana, I. G. A. G., & Liestiandre, H. K. (2021). COVID-19 perceptions, subjective norms, and perceived benefits to attitude and behavior of continuous using of medical mask. *Linguistics and Culture Review*, *5*(S2), 1259–1280. https://doi.org/10.21744/lingcure.v5ns2.1805

Yoon, J. H., & Cho, E. (2021). Effectuation (EF) and Causation (CS) on Venture Performance and Entrepreneurs' Dispositions Affecting the Reliance on EF and CS. *Entrepreneurship Research Journal*, 20200054. https://doi.org/doi:10.1515/erj-2020-0054

Zhang, S. N., Li, Y. Q., Liu, C. H., & Ruan, W. Q. (2020). Critical factors identification and prediction of tourism and hospitality students' entrepreneurial intention. *Journal of Hospitality, Leisure, Sport and Tourism Education*, *26*(December 2019), 100234. https://doi.org/10.1016/j.jhlste.2019.100234